\documentclass{revtex4}

\usepackage{graphicx}
\def\mz{m_Z}

\def\hpm{H^{\pm}}

\def\tauptaum{\tau^+\tau^-}

\def\lsim{\mathrel{\raise.3ex\hbox{$<$\kern-.75em\lower1ex\hbox{$\sim$}}}}
\def\gsim{\mathrel{\raise.3ex\hbox{$>$\kern-.75em\lower1ex\hbox{$\sim$}}}}
\def\ifmath#1{\relax\ifmmode #1\else $#1$\fi}

\def\hsm{h_{\rm SM}}
\def\mhsm{m_{\hsm}}
\def\hl{h^0}
\def\hh{H^0}
\def\ha{A^0}

\def\hpm{H^{\pm}}
\def\mhl{m_{\hl}}
\def\mhh{m_{\hh}}
\def\mha{m_{\ha}}

\def\mhpm{m_{\hpm}}
\def\tanb{\tan\beta}

\def\mz{m_Z}
\def\mw{m_W}

\def\wp{W^+}
\def\wm{W^-}

\def\MPL #1 #2 #3 {{\sl Mod.~Phys.~Lett.}~{\bf#1} (#3) #2}
\def\NPB #1 #2 #3 {{\sl Nucl.~Phys.}~{\bf #1} (#3) #2}
\def\PLB #1 #2 #3 {{\sl Phys.~Lett.}~{\bf #1} (#3) #2}
\def\PR #1 #2 #3 {{\sl Phys.~Rep.}~{\bf#1} (#3) #2}
\def\PRD #1 #2 #3 {{\sl Phys.~Rev.}~{\bf #1} (#3) #2}
\def\PRL #1 #2 #3 {{\sl Phys.~Rev.~Lett.}~{\bf#1} (#3) #2}
\def\RMP #1 #2 #3 {{\sl Rev.~Mod.~Phys.}~{\bf#1} (#3) #2}
\def\ZPC #1 #2 #3 {{\sl Z.~Phys.}~{\bf #1} (#3) #2}
\def\IJMP #1 #2 #3 {{\sl Int.~J.~Mod.~Phys.}~{\bf#1} (#3) #2}
\def\NIM #1 #2 #3 {{\sl Nucl.~Inst.~and~Meth.}~{\bf#1} {#3} #2}

\def\tauptaum{\tau^+\tau^-}

\def\gam{\gamma}

\def\anti{\overline}
\def\epem{e^+e^-}

\def\rts{\sqrt s}
\def\ie{{\it i.e.}}

\def\anti{\overline}

\def\fbi{~{\rm fb}^{-1}}
\def\fb{~{\rm fb}}

\def\mev{~{\rm MeV}}
\def\gev{~{\rm GeV}}
\def\tev{~{\rm TeV}}

\newcommand{\nc}{\newcommand}
\nc{\beq}{\begin{equation}}   \nc{\eeq}{\end{equation}}
\nc{\bea}{\begin{eqnarray}}   \nc{\eea}{\end{eqnarray}}
\nc{\baa}{\begin{array}}      \nc{\eaa}{\end{array}}
\nc{\bit}{\begin{itemize}}    \nc{\eit}{\end{itemize}}
\nc{\ben}{\begin{enumerate}}  \nc{\een}{\end{enumerate}}
\nc{\bce}{\begin{center}}     \nc{\ece}{\end{center}}
\def\beqa{\begin{eqnarray}}
\def\eeqa{\end{eqnarray}}
\def\bed{\begin{description}}
\def\eed{\end{description}}

\def\tanb{\tan\beta}

\begin{document}
\bibliographystyle{revtex}

\begin{titlepage}
\def\thefootnote{\fnsymbol{footnote}}       

\begin{center}
\mbox{ } 

\end{center}
\vskip 1.0cm

\begin{flushright}
\Large
\mbox{\hspace{10.2cm} hep-ph/0202087} \\
\mbox{\hspace{12.0cm} February, 2002}
\end{flushright}
\begin{center}
\vskip 3.5cm
{\LARGE\bf Higgs Sectors in which the only light Higgs boson is \\[5pt]CP-odd 
and Linear Collider Strategies for its Discovery
}
\vskip 1cm
{{\LARGE\bf Tom Farris$^1$, John F. Gunion$^1$, Heather E. Logan$^2$}\\
\smallskip
\Large 
$^1$Davis Institute for HEP, U. of California, Davis, CA \\
$^2$Fermilab, Batavia, IL }

\vskip 2.5cm
\centerline{\Large \bf Abstract}
\end{center}

\begin{center}
\begin{minipage}{16cm}
\large
We survey techniques for finding
a CP-odd Higgs boson, $\ha$, at the Linear Collider 
that do not depend upon the presence of other light Higgs bosons.
The potential reach in $[\mha,\tanb]$ parameter space for
various production/discovery modes is evaluated 
and regions where discovery might not be possible 
at a given $\rts$ are delineated. We give, for the first time,
results for $\epem\to \nu\anti\nu \ha$ one-loop $W$ boson fusion production.

\vspace{1cm}

{\sl 
Contribution to the 
Snowmass 2001 Workshop on ``The Future of Particle Physics'', Snowmass, CO, USA, July 2001
\vspace{-3cm}
}

\end{minipage}
\end{center}

\end{titlepage}

\newpage

\title{Higgs Sectors in which the only light
Higgs boson is CP-odd \\ and Linear Collider Strategies for its Discovery}



\author{Tom Farris$^1$, John F. Gunion$^1$ and Heather E. Logan$^2$}
\affiliation{$^1$Department of Physics, University of California, Davis, CA 95616\\ $^2$Fermilab, Batavia, IL 60510}



\begin{abstract}
We survey techniques for finding
a CP-odd Higgs boson, $\ha$, at the Linear Collider 
that do not depend upon the presence of other light Higgs bosons.
The potential reach in $[\mha,\tanb]$ parameter space for
various production/discovery modes is evaluated 
and regions where discovery might not be possible 
at a given $\rts$ are delineated. We give, for the first time,
results for $\epem\to \nu\anti\nu \ha$ one-loop $W$ boson fusion production.

\end{abstract}

\maketitle

\noindent 
A general two-Higgs-doublet model (2HDM) or more complicated
extension of the one-doublet Higgs sector of the Standard Model (SM)
remains an attractive possibility~\cite{hhg}, especially as an effective
theory in the context of models with new physics at an energy scale 
significantly below the usual GUT scale. Although gauge coupling
unification is not necessarily relevant in such theories, it 
can be achieved \cite{Gunion:1998ii}. For example, for two doublets
and one $T=1,Y=0$ triplet, the gauge couplings unify at 
$1.6\times 10^{14}\gev$; increasingly complicated Higgs sectors
are required for gauge coupling unification at still lower scales.
(The unification at low scales cannot be true gauge group
unification without encountering problems with proton decay. However,
there are examples of theories (for example, many string theories) in
which the couplings are predicted to unify without 
the presence of a larger gauge group.)
If there is a neutral member of a triplet representation,
$\rho=\mw/(\mz\cos\theta_W)=1$ remains natural provided
it has zero vev \cite{Gunion:1991dt}. 

Current data provide some important hints and constraints regarding
the Higgs sector \cite{lepewwg}.  As is well known, the simplest interpretation
of the precision electroweak data is the existence of a rather
light SM-like Higgs boson (the mass corresponding to the smallest
$\chi^2$ being $\sim 88\gev$,  well below the LEP
experimental lower limit of 114.1 GeV). However, alternative 
fits to the precision electroweak data without a light SM-like
Higgs boson are possible when an extended Higgs sector is present.
We will focus on the CP-conserving (CPC) 2HDM 
with its five physical Higgs bosons, $\hl,\hh,\ha,\hpm$.

\begin{figure}[h!]
\begin{center}
\includegraphics[width=12cm]{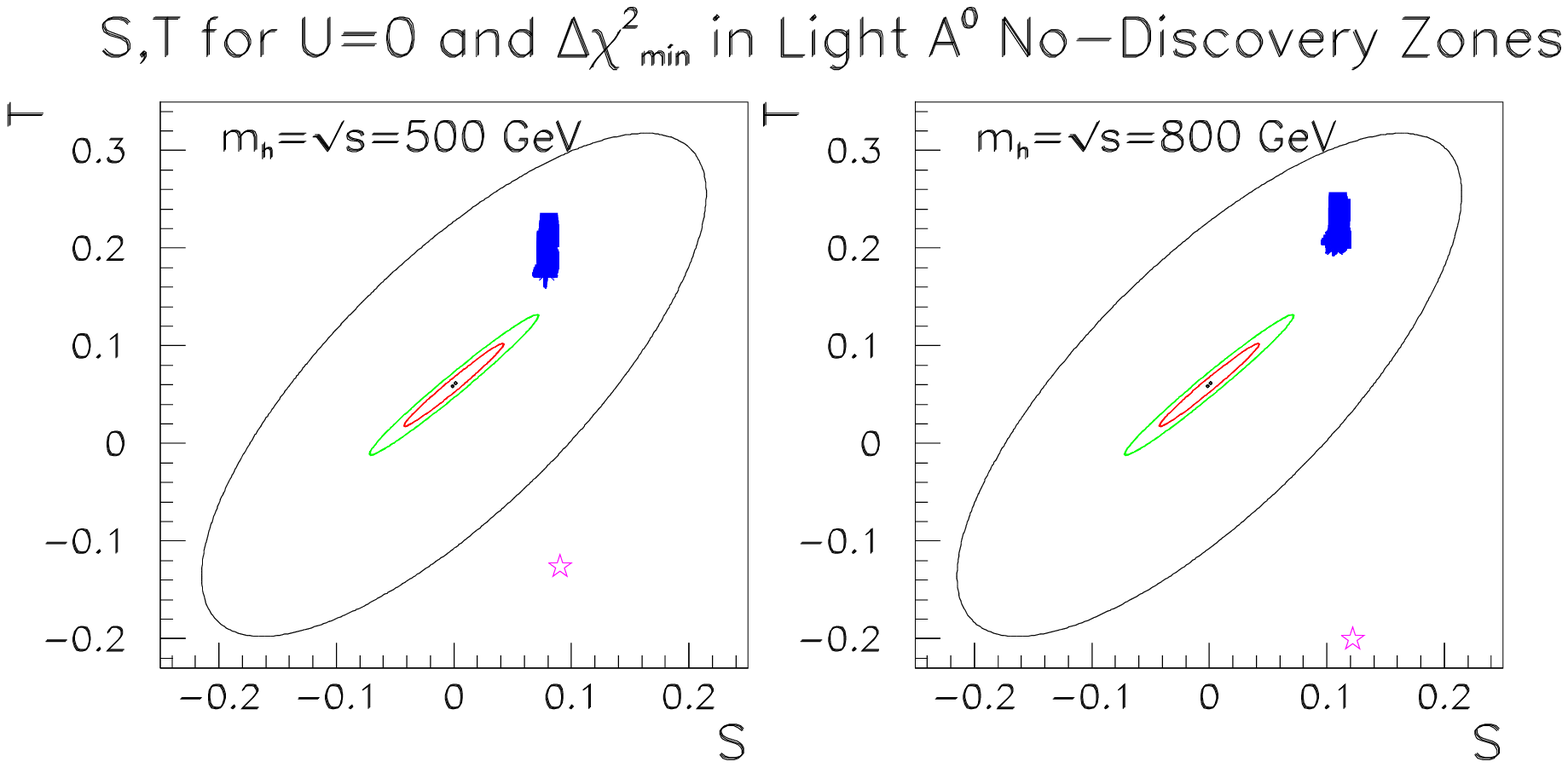}
\vspace*{-2.3in}
\caption{\label{ellipse}
The outer ellipses show the 90\% CL region from current
precision electroweak data in the $S,T$ plane
for $U=0$ relative to a central point defined
by the SM prediction with $\mhsm=115$ GeV. 
The blobs of points show the $S,T$ predictions for 2HDM  models 
with a light $\ha$ and with
$\tanb$ such that the $\ha$ cannot be detected
in $b\anti b\ha$ or $t\anti t\ha$ production
at either the LC or the LHC; the mass of
the SM-like $\hl$ is set equal to $\rts=500\gev$ (left)
or $800\gev$ (right) and $\mhpm$ and $\mhh$ have been
chosen to minimize the $\chi^2$ of the full precision electroweak fit.
The innermost (middle) ellipse shows the  90\% (99.9\%) CL
region for $\mhsm=115$ GeV after
Giga-$Z$  LC operation {\it and} a $\Delta m_W\protect\lsim 6$ MeV threshold
scan measurement. The stars to the bottom right show the $S,T$ predictions
in the case of the SM with $\mhsm=500\gev$ (left) or $800\gev$ (right).
This figure is from \cite{Gunion:2000ab}.} 
\end{center}
\end{figure}

The scenario we wish to consider is that in which
the $\ha$ of the 2HDM is light
and all other Higgs bosons are heavy.
It turns out that this type of scenario 
can be consistent with precision electroweak
constraints~\cite{Chankowski:2000an}.
If $\mha$ is small, the best fit to the precision
electroweak data is achieved by choosing 
the lighter CP-even Higgs boson, $\hl$,
to be SM-like. A good fit is achieved even
for $\mhl\sim 1\tev$. Of course, 
such a heavy SM-like $\hl$ leads to large 
$\Delta S>0$ and large $\Delta T<0$ contributions, which 
on their own would place the $S,T$ prediction of the 2HDM model 
well outside the current 90\% CL ellipse --- see the stars in 
Fig.~\ref{ellipse}  (from \cite{Gunion:2000ab}).
However, the large $\Delta T<0$ contribution from the SM-like $\hl$
can be compensated by a large $\Delta T>0$ from a
small mass non-degeneracy (weak isospin breaking) of the
still heavier $\hh$ and $\hpm$ Higgs bosons. In detail, for a light $\ha$
and SM-like $\hl$ one finds
\beq
   \Delta \rho=\frac{\alpha}{16 \pi m_W^2 c_W^2}\left\{\frac{c_W^2}{s_W^2}
   \frac{m_{H^\pm}^2-m_{H^0}^2}{2}-3m_W^2\left[\log\frac{m_{h^0}^2}{m_W^2}
   +\frac{1}{6}+\frac{1}{s_W^2}\log\frac{m_W^2}{m_Z^2}\right]\right\}\nonumber
\label{drhonew}
\eeq
from which we see that the first term can easily compensate
the large negative contribution to $\Delta\rho$ from the $\log (\mhl^2/\mw^2)$
term. In Fig.~\ref{ellipse}, 
the blobs correspond to 2HDM parameter choices for which:
(a) $\mhl=\rts$ of a linear $\epem$ collider (LC) (\ie\ $\mhl$
is such that the $\hl$ cannot be observed at the LC); (b)
$\mhpm-\mhh\sim {\rm few}\gev$ has been chosen
(with both $\mhpm,\mhh\gsim 1\tev$)
so that the $S,T$ prediction is well within the 90\% CL ellipse
of the current precision electroweak fits; and (c)
$\mha$ and $\tanb$ are in the `wedge' of $[\mha,\tanb]$ parameter space
for which detection of the $\ha$ via $t\anti t\ha$ and $b\anti b\ha$
production at the LHC and LC
would be difficult \cite{Grzadkowski:2000wj}.
(This wedge will be discussed in more detail below.
For $\rts=1\tev$ and $L=1000\fbi$ at the LC, $\mha$ values as low as
roughly $100\gev$ could still fall into this wedge for $\tanb\sim 5$.) 
However, this scenario
can only be pushed so far. In order to maintain perturbativity
for all the Higgs self couplings, it is necessary that the $\hl$, $\hh$
and $\hpm$ masses not be greatly in excess of $1\tev$. This implies, in
particular, that the SM-like $\hl$ would be detected at the LHC.
If it should happen that a heavy SM-like Higgs boson is detected
at the LHC, the precision electroweak situation could
only be resolved by Giga-$Z$ operation 
and a $\Delta\mw=6\mev$ $WW$ threshold scan at the LC
(with the resulting ellipse sizes illustrated in Fig.~\ref{ellipse}).
The resulting determination of $S,T$ 
would be sufficiently precise to definitively check for values
like those of the blobs of Fig.~\ref{ellipse}. If no
other new physics was detected at the LC or LHC that could cause
the extra $\Delta T>0$, searching for the other Higgs
bosons of a possible 2HDM Higgs sector, 
especially a possibly light decoupled $\ha$, 
would become a high priority.

Interestingly, a light $\ha$ with $\mha\gsim 10\gev$
would yield a positive contribution to 
the muon's anomalous magnetic moment, $a_\mu$, coming mainly from the two-loop
Bar-Zee graph  \cite{Cheung:2001hz,Krawczyk:2001nw}. Recent 
data \cite{Brown:2001mg} suggests the presence of just such a deviation
from SM expectations.
However, after including the recent corrections
to the sign of the light-by-light scattering contribution and
allowing for uncertainty in $\sigma_{\rm had}$, the discrepancy 
between the experimental result and the SM prediction is not large,
and may not be present at all. The best that can be said is that a small 
positive discrepancy in $a_\mu$ can be explained by the presence of an $\ha$
for moderate values of $\mha$ and $\tanb$ in the `wedge' region
for which direct discovery of the $\ha$ would be difficult at the LC and LHC
using the standard modes we describe later.

To summarize, it is not unreasonable to suppose that the Higgs
sector contains a 2HDM with a light $\ha$, a heavy SM-like $\hl$,
and still heavier $\hh$ and $\hpm$ with a small mass splitting.  
The $\hl$ would be
detected at the LHC, but we would have no understanding of
how this is to be made consistent with precision electroweak constraints.
Direct detection of the $\ha$ would become a priority.
In the remainder of this note, we wish to consider the various
means for detecting a light $\ha$ at a linear collider.
\begin{figure}[h]
\begin{center}
\vspace*{-.2in}
\begin{minipage}{7.2in}
\hskip -.3in
\includegraphics[width=3.5in]{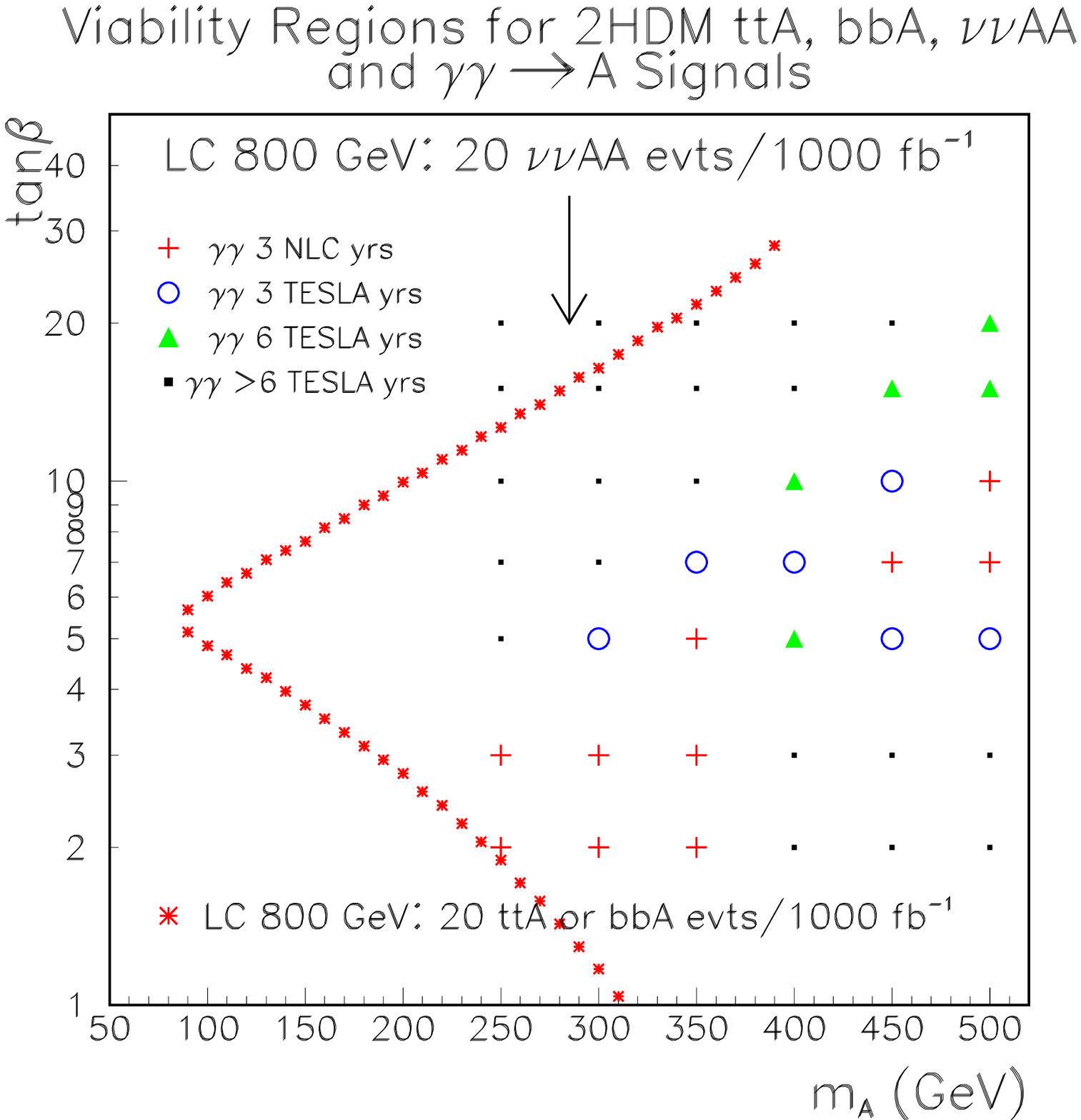}\hspace*{-.25in}\includegraphics[width=3.7in]{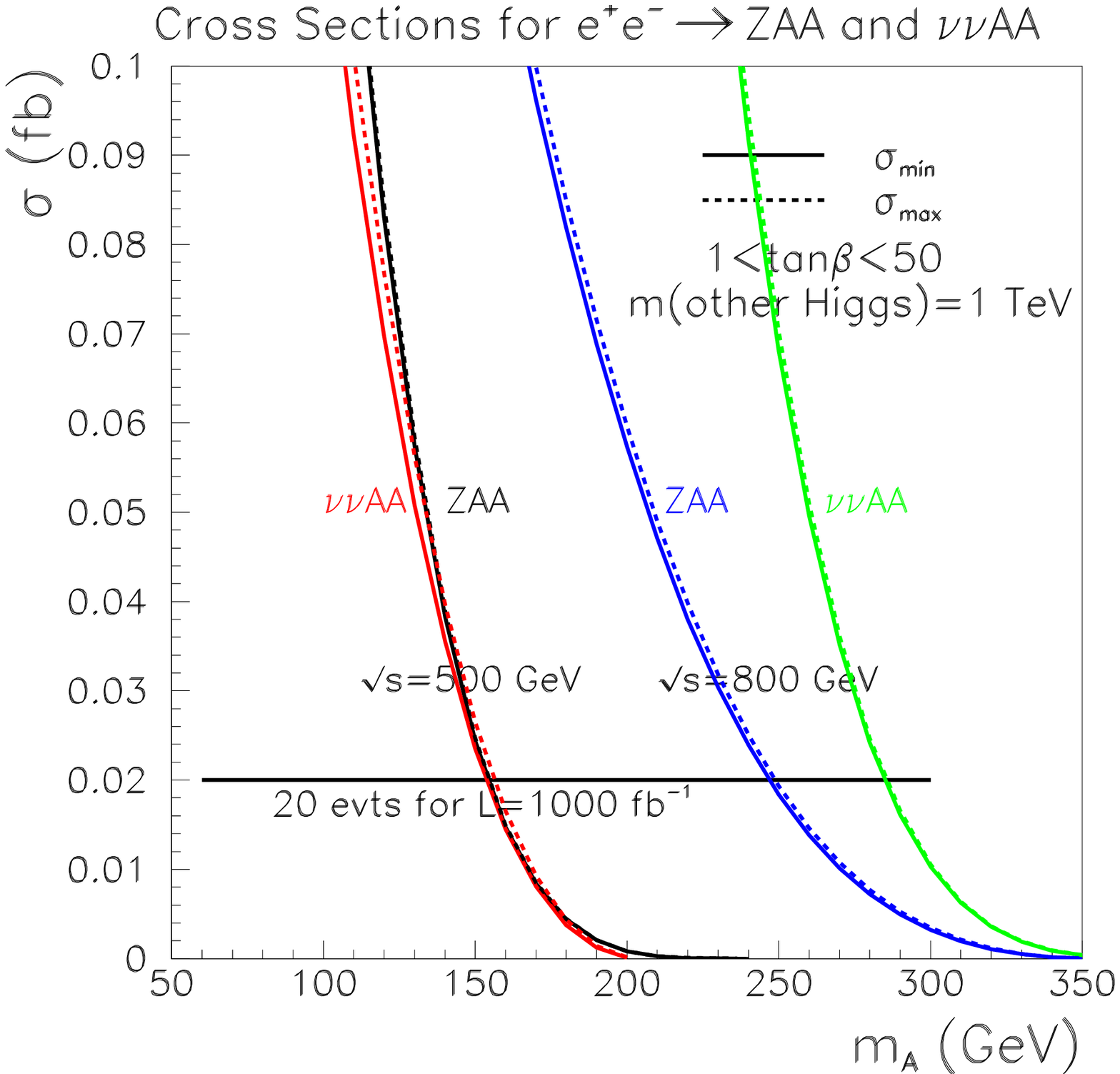}
\end{minipage}
\vspace*{.4in}
\begin{minipage}{3.in}
\hspace*{.2in}Figure 2
\end{minipage}
\begin{minipage}{3.5in}
\hskip .5in
Figure 3
\end{minipage}
\vspace*{-.5in}
\caption{\label{wedge} We display (using stars) the wedge of $[\mha,\tanb]$
parameter space inside which a LC operating at $\rts=800\gev$ yields
fewer than 20 events per $1000\fbi$ 
in both the $t\anti t \ha$ and $b\anti b \ha$
production modes. Also shown by the arrow is the $\mha$ value above
which the process $\epem\to \nu\anti\nu \ha\ha$ yields fewer than 20
events per $1000\fbi$. The $+$ symbols on the grid of $[\mha,\tanb]$ values
show the points for which a $4\sigma$ signal for $\gam\gam\to\ha$
would be achieved using NLC operation at $\rts=630\gev$ after three $10^7$ sec 
years of operation assuming running conditions 
and strategies as specified in \cite{gunasner}.
In particular, operation for two years using a polarization
configuration for the electron beams and laser photons 
yielding a broad $E_{\gam\gam}$ spectrum
and one year using a configuration yielding a peaked spectrum is assumed. 
The circles (triangles) show the additional points that would yield
a $4\sigma$ signal for twice (four times)
the integrated luminosity of the current
NLC $\gam\gam$ interaction region design. The factor of two increase 
is probably attainable at TESLA.
The small squares show the additional points
sampled in the study of \cite{gunasner}.}
\caption[0]{\label{aaz_nnaa_fig} 
We plot the cross sections for $\epem\to Z\ha\ha$ and $\epem\to \nu\anti\nu
\ha\ha$ as a function of $\mha$, assuming a 2HDM model with a heavy
SM-like $\hl$. We have taken $\mhl=\mhh=\mhpm=1\tev$.
Maximum and minimum values found after scanning $1\leq\tanb\leq 50$
are shown for $\rts=500\gev$ and $800\gev$.  
(The variation with $\tanb$ arises from small contributions
associated with exchanges of the heavy Higgs bosons.) The 20 event level
for $L=1000\fbi$ is indicated.}
\vspace*{-.35in}
\end{center}
\end{figure}

At the LC, the relevant discovery processes for a $\ha$ with no tree-level
$WW,ZZ$ couplings are: $\epem\to t\anti t \ha$ and 
$\epem\to b\anti b \ha$~\cite{Djouadi:gp,Grzadkowski:1999ye}
\cite{Grzadkowski:2000wj}; 
$\epem\to Z^*\to  Z \ha\ha$ \cite{Haber:1993jr}; 
$\epem\to\nu\anti\nu W^*W^*\to\nu\anti\nu  \ha\ha$~\cite{Djouadi:1996jf};
$\gam\gam\to \ha$~\cite{gunasner} (see also \cite{Muhlleitner:2001kw}). 
That these processes might have reasonable rates follows from
the couplings involved. At least one of the $\gamma_5$ Yukawa
couplings of the $\ha$ must be substantial: relative to SM-like strength,
$t\anti t\ha=\cot\beta$ and $b\anti b\ha=\tan\beta$.
The quartic couplings,
$ZZ\ha\ha$ and $\wp\wm\ha\ha$, arise from the gauge covariant
structure  $(D_\mu \Phi)^\dagger (D^\mu \Phi)$ and are 
of guaranteed magnitude. The
$\gam\gam\to\ha$ coupling derives from fermion loops, and, as noted,
not both the $b\anti b\ha$ and $ t\anti t \ha$ coupling
can be suppressed.

Turning first to $t\anti t\ha$ and $b\anti b\ha$ production,
the former (latter) always yields significant rates if $\tanb$ is small (large)
enough (and the process is kinematically allowed).
But, even for high $\rts$ and large luminosity,
there remains a wedge of moderate $\tanb$ for which 
neither process provides adequate event rate \cite{Grzadkowski:2000wj,Grzadkowski:1999ye}.
The wedge corresponding to fewer than 20 events in either process
for $L=1000\fbi$ at $\rts=800\gev$
is shown in Fig.~\ref{wedge}. (Probably backgrounds
would imply that more than 20 events would be needed to see the signal,
so this wedge is a conservative indication of the region
in which these processes would not be visible.)
The extent of the corresponding wedge at the LHC
can be estimated from the CMS and ATLAS \cite{1r,2r}
$[\mha,\tanb]$ discovery region plots for the MSSM Higgs sector as follows.
At high $\tanb$, the $b\anti b\hh$ and $b\anti b\ha$ 
processes make roughly equal contributions
to the $b\anti b\tauptaum$ final state signal.  
Since the rates are proportional to $\tan^2\beta$,
the location of the upper limit of the LHC wedge simply needs to be
rescaled by a factor of $\sqrt 2$, implying the LHC could find a $\ha$
signal for $\tanb>14$ at $\mha=250\gev$ 
(comparable to the LC result) rising to $\tanb>24$ at $\mha=500\gev$
(which is significantly better coverage than the LC).
However, at low $\tanb$, the only MSSM channel for 
$\ha$ discovery at the LHC deemed viable to date employs $\ha\to Z\hl$
decays which would be absent in the type of model being
considered here in which only the $\ha$ is light.

For the lower values of $\mha$, double Higgs production via the quartic
couplings will allow discovery at the LC even in the wedge region.
The cross sections for $\epem\to Z^*\to Z\ha\ha$ 
and $\epem\to \nu\anti\nu\ha\ha$ are shown in Fig.~\ref{aaz_nnaa_fig}.
For instance, the process $\epem\to Z^*\to Z\ha\ha$ yields
20 events for $L=1000\fbi$ for 
$\mha\lsim 160\gev$ ($\mha\lsim 250\gev$) for $\rts=500\gev$ ($\rts=800\gev$),
while $WW\to \ha\ha$ fusion production yields 20 events for 
$\mha\lsim 160\gev$ ($\mha\lsim 290\gev$), respectively.
A careful assessment of backgrounds is required to ascertain just
what the mass reach of these processes actually is.

If the $\gam\gam$ collider option is implemented at the LC,
$\gam\gam\to \ha$ will provide a signal for a decoupled $\ha$
over a significant portion of the wedge region. The
results from the quite realistic study of \cite{gunasner} are
illustrated in Fig.~\ref{wedge}, which focuses on
$\mha\geq 250\gev$. The 
pluses indicate $4\sigma$ discovery points after 3 years of appropriate
running at the NLC. The higher TESLA luminosity for $\gam\gam$ collisions
would allow $4\sigma$ discovery for the additional points indicated
by the circles.

Finally, although we don't present details here, a muon collider 
capable of operating at $\rts=500\gev$ and below would
probably be able to provide $4\sigma$ signals for the $\ha$ in the 
$\mha<500\gev$
wedge region after about 3 years of appropriately configured operation,
assuming the current nominal Higgs factory luminosities.
For more details, see \cite{jfgucla}.

The above results indicate the need for exploring additional
mechanisms by which a $\ha$ with $\mha\geq 250\gev$ might be
produced and detected.  The remaining possibilities are the one-loop
processes: $\epem\to \gam\ha$, $\epem\to Z\ha$ and $\epem\to \nu\anti\nu\ha$.
The first two have previously been explored 
in \cite{Djouadi:1996ws,Akeroyd:1999gu}. Results for
the third process will be given for the first time here; details 
of the computation will appear in \cite{nnaprocess}. The results 
we shall present for $\epem\to\gam\ha$ agree 
(where comparison is possible) with the 2HDM results of \cite{Djouadi:1996ws},
but not with those of \cite{Akeroyd:1999gu}.  Our results for $\epem\to Z\ha$
do not agree except in a very rough way with the 2HDM
results of \cite{Akeroyd:1999gu}.
(The $\epem\to Z\ha$ process was not computed in \cite{Djouadi:1996ws}.)
In all our loop computations, we have employed the running $b$-quark
mass as a function of $\mha$ in evaluating the $b\anti b\ha$
coupling employed in computing the $b$-quark loop contribution
to the one-loop couplings.  Our results are obtained by including
only the fermion $b,t$ loop contributions; in particular,
we assume that all other Higgs bosons are sufficiently heavy that
loop diagrams containing them will be small.

\begin{figure}[p]
\begin{center}
\includegraphics[width=7.5cm,angle=90]{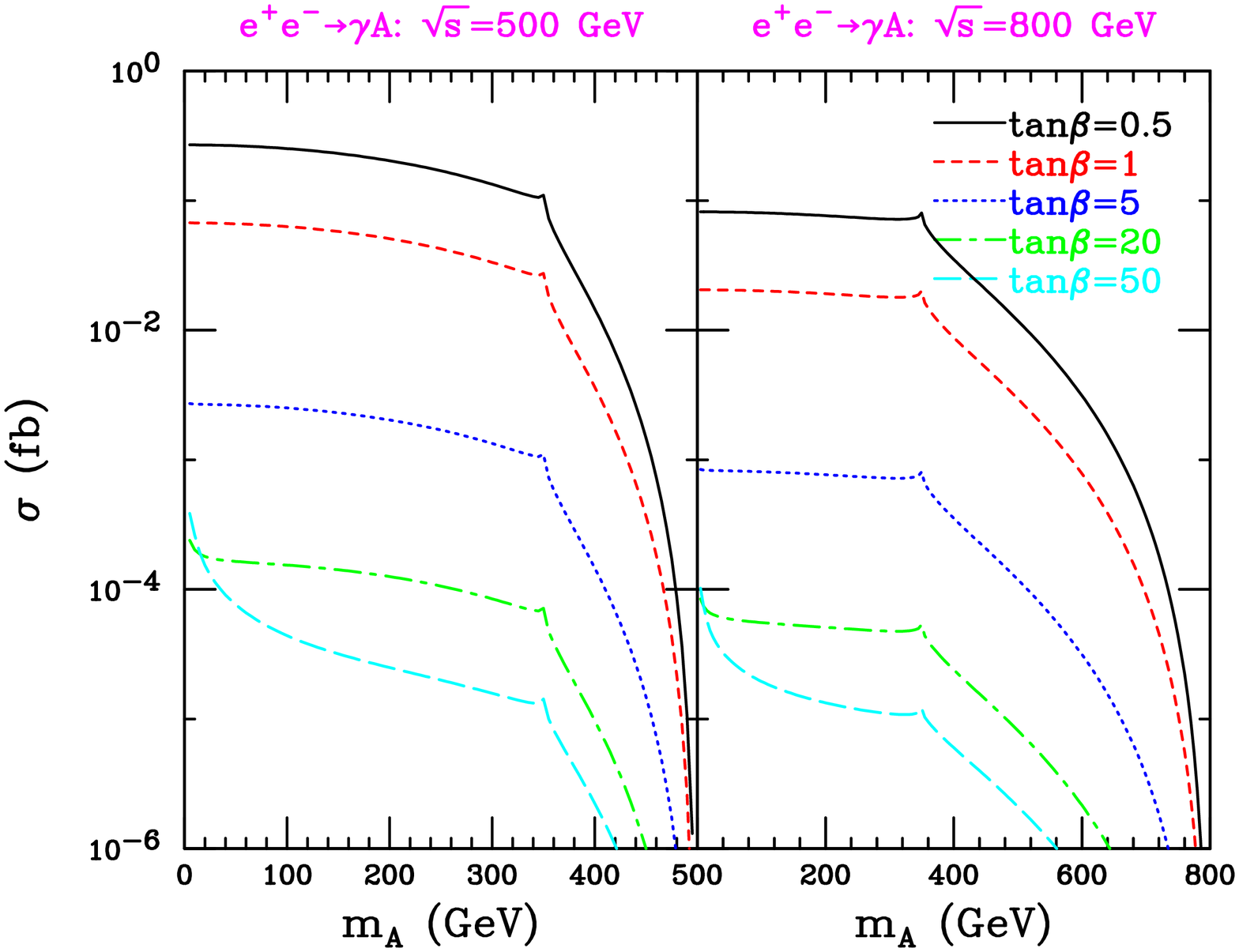}
\includegraphics[width=7.5cm,angle=90]{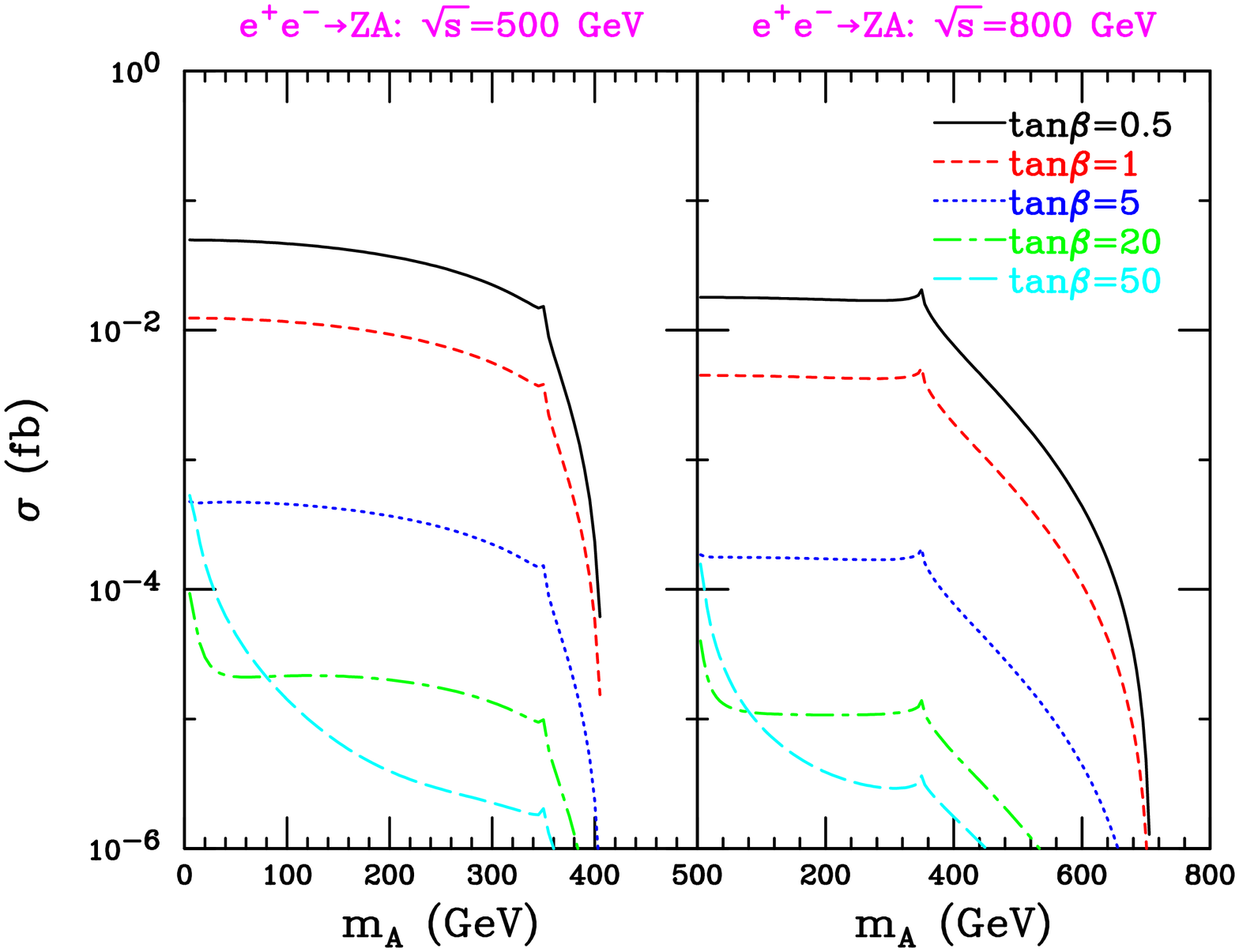}
\includegraphics[width=7.5cm,angle=90]{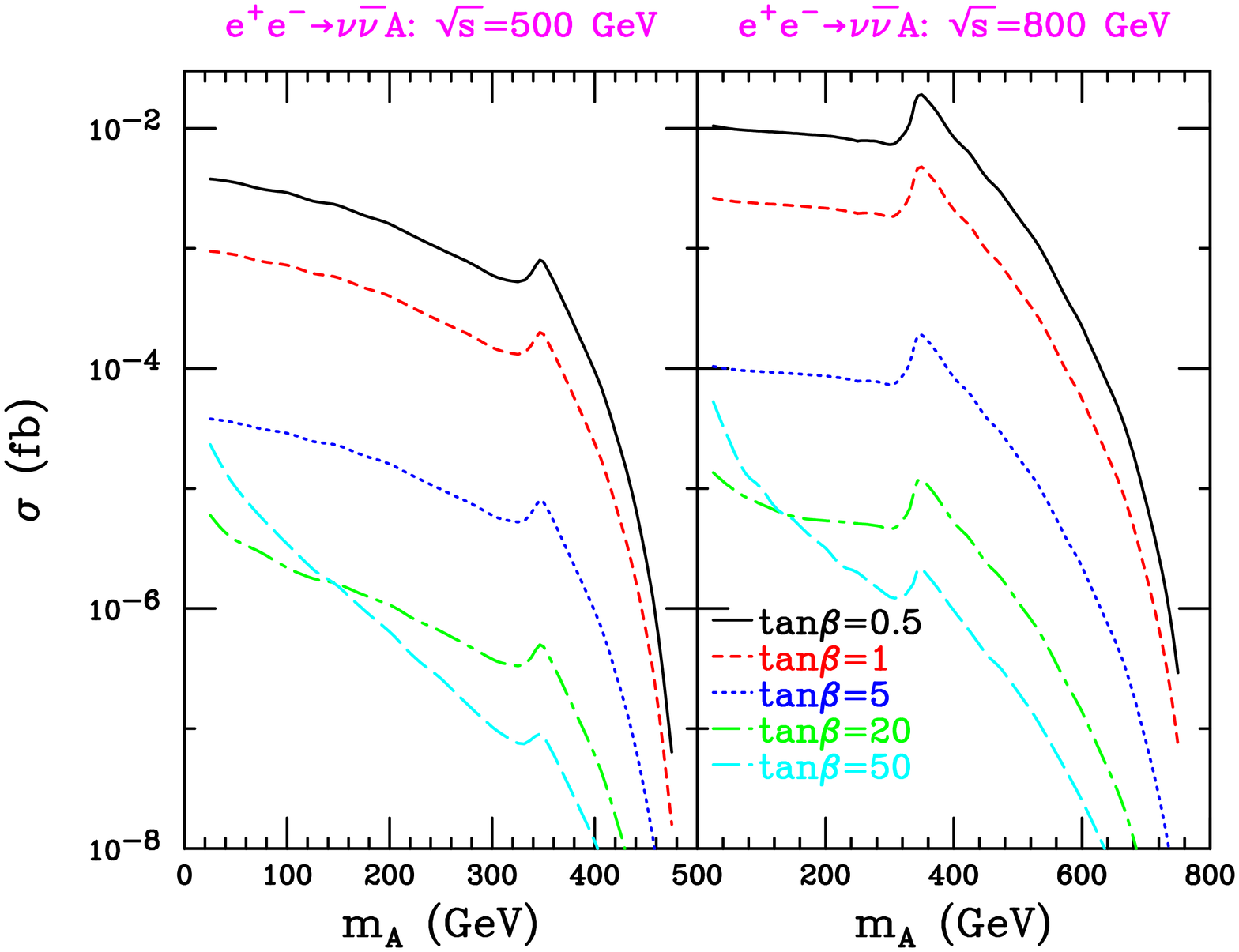}
\caption{\label{loopha} The $\epem\to\gam\ha$, $Z\ha$ and $\nu\anti\nu\ha$ 
cross sections as a function of $\mha$ for $\rts=500\gev$ and $800\gev$,
for $\tanb=0.5,1,5,20,50$.}
\end{center}
\end{figure}
Our results for $\epem\to \gam\ha$, $Z\ha$ and $\nu\anti\nu\ha$
appear in the three windows of Fig.~\ref{loopha}.
From these figures, it should be immediately clear that
the only process that might yield a useful event rate is
$\epem\to\gam\ha$, and then only if $\tanb$ is not large. 
For $\epem\to\gam\ha$, roughly 80 declining to 30 events
are predicted for $\tanb=1$ and $\mha=20\gev$ increasing to $350\gev$,
assuming $\rts=500\gev$ and $L=1000\fbi$. At 
$\tanb=5$, only 3 declining to 1 events
are anticipated for the same $\mha$ mass range. Unfortunately, there
will be substantial background.  Assuming that the search will
take place in the $\gam b\anti b$ final state, the irreducible
background will come from $\epem\to\gam b\anti b$ production.
This was evaluated in \cite{Casalbuoni:1998fs}. The result found is
$d\sigma/dm_{b\anti b}=0.5 \fb/(10\gev)$ [$0.2\fb/(10\gev)$]
at $m_{b\anti b}=200\gev$ [$400\gev$] at $\rts=500\gev$. Even
if an optimistic mass resolution of 
$\Delta m_{b\anti b}=5\gev$ 
can be achieved, we see that this irreducible background
will be at the level of 
250 to 100 events in the indicated mass range. In addition,
other backgrounds as well as efficiencies for tagging 
and event selection must be taken into account.  Thus, our conclusion
is that the one-loop processes are unlikely to provide a measurable
signal, and certainly cannot be used as discovery modes, unless
$\tanb <1$.

In conclusion,
there are a variety of perfectly viable Higgs sector models
in which it would be highly desirable
to be able to detect a relatively light CP-odd $\ha$ 
without relying on associated production with other Higgs bosons.
Such detection might be crucial
to determining the nature of the Higgs sector but may be quite difficult.
A linear $\epem$ collider, including the $\gam\gam$ collider option,
provides the best range of possibilities for $\ha$ discovery. 
Even when the $\epem\to \nu\anti\nu \ha\ha$
pair process becomes strongly kinematically suppressed
(roughly $\rts<200\gev+2\mha$), $\gam\gam\to\ha$ production continues to 
provide an opportunity for $\ha$ discovery in the moderate-$\tanb$ 
`wedge' region of $[\mha,\tanb]$ parameter space where $t\anti t\ha$
and $b\anti b \ha$ production both fail.  Although this will still
leave some portions of $[\mha,\tanb]$ parameter space inaccessible
to $\ha$ discovery, it is quite impressive that
the tools and techniques 
that have been developed for Higgs detection at the LC
have reached a high enough
level of sophistication that we should have
a good chance of detecting and studying
the Higgs bosons of even rather unusual Higgs sectors.

\vspace*{-.3in}
\begin{acknowledgements}
{\vspace*{-.2in}TF and JFG are 
supported in part by the U.S. Department
of Energy and by the Davis Institute for High Energy Physics.
Fermilab is operated by Universities Research Association Inc.\ 
under contract no.~DE-AC02-76CH03000 with the U.S. Department of
Energy. We wish to thank J. Kalinowski and S. Su for helpful
comments and conversations.
}
\end{acknowledgements}
\vspace*{-.25in}

%
%

%
%



\end{document}